\documentclass{bioinfo}
\copyrightyear{2014}
\pubyear{2014}
\usepackage{cite} 
\usepackage{url}  
\usepackage{booktabs}
\usepackage{ifpdf}

\newcounter{lnoc}
\newenvironment{myalgorithm}[1]{%
\hrule height 0.8pt \vspace{0.6ex} \footnotesize#1\vspace{0.6ex}\hrule height 0.5pt \vspace{-1.5ex}
\setcounter{lnoc}{0}
\scriptsize
\begin{tabbing}
00000\=XXI\=XXI\=XXI\=XXI\=XXI\=XXI\=\kill
}{%
\end{tabbing}
\vspace{-2.0ex}\hrule height 0.8pt}
\newcommand{\lno}[1][]{{\scriptsize\sffamily \stepcounter{lnoc} 
\ifnum\thelnoc<10
\phantom0%
\fi
\thelnoc#1}\>}

\newcommand{\pcfor}{{\bfseries for~}}
\newcommand{\pcto}{{\bfseries to~}}

\newcommand{\pcdo}{{\bfseries do~}}
\newcommand{\pcif}{{\bfseries if~}}
\newcommand{\pcthen}{{\bfseries then~}}
\newcommand{\pcelse}{{\bfseries else~}}
\newcommand{\pcelseif}{{\bfseries else if~}}
\newcommand{\pcwhile}{{\bfseries while~}}

\newcommand{\pcfun}[1]{\mbox{\textrm{#1}}}
\newcommand{\pccomment}[1]{\qqq\{\textit{#1}\}}

\newcommand{\pcfalse}{\mbox{\textrm{\bfseries  false}}}

\newcommand{\band}{\mbox{\textsf{\&}~}}

\newcommand{\bneg}{\mbox{\textsf{\~{ }}}}

\newcommand{\tuple}[1]{$\langle\mathit{#1}\rangle$}

\usepackage{ifpdf}

\newcommand{\q}{\phantom0}
\newcommand{\qq}{\phantom{00}}
\newcommand{\qqq}{\phantom{000}}

\newcommand{\qc}{\phantom{0,}}

\begin{document}
\firstpage{1}
\title{Indexing large genome collections on a PC}
\author[Agnieszka Danek, Sebastian Deorowicz, Szymon Grabowski]{Agnieszka Danek\,$^{1}$, Sebastian Deorowicz\,$^{1}\footnote{To whom correspondence should be addressed}$, Szymon Grabowski\,$^{2}$}
\address{
$^{1}$Institute of Informatics, Silesian University of Technology, Akademicka 16, 44-100 Gliwice, Poland\\
$^{2}$Computer Engineering Department, Technical University of {\L}\'{o}d\'{z}, Al.\ Politechniki 11, 90-924 {\L}\'{o}d\'{z}, Poland}

\history{Received on XXXXX; revised on XXXXX; accepted on XXXXX}
\editor{Associate Editor: XXXXXXX}
\maketitle

\begin{abstract}
\section{Motivation:}
The availability of thousands of invidual genomes of one species
should boost rapid progress in personalized medicine or
understanding of the interaction between genotype and phenotype,
to name a few applications.
A key operation useful in such analyses is aligning sequencing reads
against a collection of genomes, which is costly with the use of
existing algorithms due to their large memory requirements. 

\section{Results:}
We present MuGI, Multiple Genome Index, which reports
all occurrences of a given pattern, in exact and approximate matching model,
against a collection of thousand(s) genomes.
Its unique feature is the small index size fitting in a standard computer
with 16--32\,GB, or even 8\,GB, of RAM, for the 1000GP collection
of 1092 diploid human genomes.
The solution is also fast.
For example, the exact matching queries are handled in average
time of 39\,$\mu$s and with up to 3 mismatches in 373\,$\mu$s
on the test PC with the index size of 13.4\,GB.
For a smaller index, occupying 7.4\,GB in memory,
the respective times grow to 76\,$\mu$s and 917\,$\mu$s. 

\section{Availability:}
Software and Supplementary material:\newline \url{http://sun.aei.polsl.pl/mugi}.


\section{Contact:} \href{sebastian.deorowicz@polsl.pl}{sebastian.deorowicz@polsl.pl}

\end{abstract}

\section{Introduction}
About a decade ago, thanks to breakthrough ideas in succinct indexing 
data structures, it was made clear that a full mammalian-sized genome 
can be 
stored and used in indexed form in main memory of a commodity workstation 
(equipped with, e.g., 4\,GB of RAM).
Probably the earliest such attempt, 
by Sadakane and Shibuya~(\citeyear{SS2001}), 
resulted in approximately 2\,GB sized compressed suffix array 
built for the April 2001 draft assembly by Human Genome Project 
at UCSC.\footnote{Obtaining low construction space, however, 
was more challenging, although later more memory frugal (or disk-based) 
algorithms for building compressed indexes appeared, 
see, e.g.,~\citep{HSS2009} and references therein.}
Yet around 2008, 
only a few sequenced human genomes were available, 
so the possibility to look for exact or approximate occurrences of a given 
DNA string in a (single) genome was clearly useful.
Nowadays, when repositories with a thousand or more genomes 
are easily available, the life scientists' goals are also more ambitious, 
and it is desirable to search for patterns in large genomic collections.
One application of such a solution could be 
simultaneous alignment of sequencing reads against multiple genomes~\citep{SHOWGKW2009}.

Interestingly, this is a largely unexplored area yet.
On one hand, toward the end of the previous decade it was noticed that the 
``standard'' compressed indexes (surveyed in~\citep{NM2007}),
e.g. from the FM or CSA family, 
are rather inappropriate to handle large collections of genomes 
of the same species, because they cannot exploit well the specific 
repetitiveness.
On a related note,
standard compression methods were inefficient 
for a simpler problem of merely compressing multiple genomes. 
Since around 2009 we can observe a surge of interest in practical, 
multi-sequence oriented DNA compressors~\citep{CLLX2009,BWB2009,CFMPN2010,KPZ2010,KPZ2011a,DG2011b,KN2013,YWLWX2013,PWY2013,DDG2013,WL2013}, 
often coupled with random access capabilities
and sometimes also offering indexed search.
The first algorithms from 2009
were soon followed 
by more mature proposals, which will be presented below, focusing on their indexing capabilities.
More information on genome data compressors and indexes can be found 
in the recent surveys~\citep{VBFD2012,DG2013,GRU2013}.

M\"akinen {\it et~al.}~(\citeyear{MNSV2010}) added index functionalities to compressed
DNA sequences: 
\emph{display} 
(which can also be called the random access functionality) 
returning the substring specified by its start and end position, 
\emph{count} 
telling the number of times the given pattern occurs in the text, 
and \emph{locate}
listing the positions of the pattern in the text.
Although those operations are not new in full-text indexes
(possibly also compressed), the authors noticed that the existing
general solutions,
paying no attention to long repeats in the input,
are not very effective here and they proposed novel {\em self-indexes}
for the considered problem.

Claude {\it et~al.}~(\citeyear{CFMPN2010}) pointed out that the full-text indexes
from~\citep{MNSV2010}, albeit fast in counting, are 
rather slow in extracting the match locations,
a feature shared by all compressed indexes based on 
the Burrows--Wheeler transform (BWT)~\citep{NM2007}.
They proposed two schemes, one basically 
an inverted index on $q$-grams, the other being 
a grammar-based self-index.
The inverted index offers interesting space-time tradeoffs 
(on real data, not in the worst case), but can basically work with 
substrings of fixed length $q$. 
The grammar-based 
index 
is more elegant and can work with any
substring length, but uses significantly more space, is slower
and needs a large amount of RAM in the index build phase.
None of these solutions can scale to large collections of mammalian-sized genomes, 
since even for 37 sequences of S.\ cerevisiae totaling 428\,Mbases 
the index construction space is at least a few gigabytes.

While a few more indexes for repetitive data were proposed in recent 
years (e.g.,~\citep{HLSTY2010,GGP2011,DJSS2012,FGHP2013}), theoretically 
superior to the ones presented above and often handling approximate matches, 
none of them can be considered a breakthrough, at least for bioinformatics, 
since none of them was demonstrated to run on multi-gigabyte genomic data.

A more ambitious goal, of indexing 1092 
human genomes, 
was set by Wandelt {\it et~al.}~(\citeyear{WSBL2013}).
They obtained a data structure of size 115.7\,GB, 
spending 
54 hours 
on a powerful laptop.
The index (loaded to RAM for a single chromosome at a time), called RCSI,  
allows to answer exact matching queries in about 250\,$\mu$s, 
and in up to 2 orders of magnitude longer time for 
$k$-approximate matching queries, depending on the choice of $k$ (up to 5).\looseness=-1

Sir{\'e}n {\it et~al.}~(\citeyear{SVM2013}) extended the 
BWT transform 
of strings to acyclic directed labeled graphs, to support path queries as an extension to substring searching. 
This allows, e.g., for read alignment on an extended BWT index of a graph representing a {\em pan-genome}, i.e., reference genome and known variants of it.
The authors 
built an index over a reference genome and a subset of variants from 
the dbSNP database, of size less than 4\,GB and allowing to match reads 
in less than 1\,ms in the exact matching mode.
The structure, called GCSA, was built in chromosome-by-chromosome manner, 
but unfortunately, 
they were unable to finish the construction for a few ``hard'' chromosomes 
even in 1\,TB of RAM!
We also note that a pan-genome contains less information than a collection 
of genomes, since the knowledge about variant occurrences in individual 
genomes is lost.

A somewhat related work, by Huang {\it et~al.}~(\citeyear{HPB2013}), presents 
an alignment tool, BWBBLE, 
working with a multi-genome (which is basically a synonime 
to pan-genome in the terminology of~\citep{SVM2013}).
BWBBLE follows a more heuristic approach than GCSA and can be constructed 
using much more humble resources.
Its memory use, however, is over $16 n \log_2 n$ bits, where $n$ is the 
multi-genome length.
This translates to 
more than 
200\,GB of memory needed to 
build a multi-genome for a collection of 1092 human genomes.
Both BWBBLE and GCSA need 
at least 
10\,ms to find matches with up to 3 errors.

Aligning sequencing reads to a genome with possible variants 
was also recently considered in theoretical works, under the problem name of 
indexing text with wildcard positions~\citep{T2013,HKSTV2013},
where the wildcards represent SNPs.
No experimental validation of the results was presented in the cited 
papers.

Most of the listed approaches are traditional string data structures, 
in the sense that they can work with arbitrary input sequences.
The nowadays practice, however, is to represent multi-genome 
collections in repositories as basically a single reference genome, 
plus a database of possible variants (e.g., SNPs), plus information on 
which of the variants from the database actually occur in each of the 
individual genomes.
The popular VCF (Variant Call Format) format allows to keep more information 
about a sequenced 
genome than listed here, but this minimal collection representation is 
enough to export each genome to its FASTA form.
Dealing with input stored in such compact form should allow to build 
efficient indexes much more easily than following the standard ``universal'' 
way, not to say about tremendous resource savings in the index construction.\looseness=-1

This modern approach was initiated in compression-only oriented 
works~\citep{CLLX2009,PWY2013,DDG2013}, and now we propose to adapt it 
in construction of a succinct and efficient index.
According to our knowledge, this is the first full-text index capable 
to work on a scale of thousand(s) of human genomes on a PC, that is, 
a small workstation equipped with 
16--32\,GB of RAM.
What is more, for a price of some slow-down the index can be used even 
on an 8\,GB machine.
No matter the end of the space-time tradeoff we are, the index is capable 
of handling also approximate matching queries, that is, reporting patterns 
locations in particular genomes from the collection with tolerance for 
up to 5 mismatches.
As said, the index is not only compact, but also fast.
For example, if up to 3 errors are allowed, the queries are handled 
in average time of 373\,$\mu$s on the test PC 
and the index takes 13.4\,GB of memory,
or in 917\,$\mu$s when the index is of size 7.4\,GB.\looseness=-1




\begin{methods}

\section{Materials and methods}

\subsection{Datasets}

We are indexing large collection of genomes of the same species, which are represented as the reference genome in FASTA format together with the 
VCF~\citep{DAA2011} file, 
describing all possible reference sequence variations and the genotype information for each of the genome in the dataset. We are only interested in details allowing for the recovery of the DNA sequences, all non-essential fields are ignored. Therefore, the data included in the VCFmin format, 
used in~\citep{DDG2013}, are sufficient. Each line describes a possible variant that may be a single nucleotide polymorphism (SNP), a deletion (DEL), an insertion (INS) or a structural variation (SV), which is typically 
a combination
of a very long deletion and an insertion. The genotype of each genome is specified in one designated column with information if each of the variant is found in this genome. In case of diploid and phased genotypes this information concerns two basic, haploid chromosome sets for each genome and treats them independently. Thus for any phased diploid genome, its DNA sequence is twice the size the reference sequence.

In our experiments we used the data available from Phase 1 of the 1000 Genomes Project~\citep{1000GP} describing the collection of 1092 phased human genomes. We concatenated the available 24 VCF files (one for each chromosome), to get one combined VCF file, which---together with the reference sequence---is the input of our algorithm building the index.\looseness=-1

\subsection{The general idea}

Our tool, Multiple Genome Index (MuGI), performs fast approximate search for input patterns in an indexed collection of genomes of the~same species. 
The searched patterns can be provided in FASTA or FASTQ format, or as a simple list in a text file (one pattern per line). 
The index is built based on the reference genome and the VCF file describing the set. The search answers the locate query---the result consists of all positions of the pattern with respect to the reference genome along with the list of all individuals in which it can be found.

The basic search regime is exact matching.
Its enhanced version allows for searching with mismatches. 
Both modes use the seed-and-extend scheme. 
The general mechanism is to quickly find a substring of the pattern and 
then extend this seed to verify if it answers the query. 

The index has one construction-time parameter, $k$, which is the 
maximum possible length of the seed. 
The match can be found directly in the reference genome and/or in its modified form, with some of the variations introduced. 
To find the seed 
we build an array of 
all possible $k$-length sequences ($k$-mers) occurring 
in all genome sequences.
The extension step is done using the reference and the available database of variants, checking which path (that is, with which variations introduced), if any, 
allows to find the full pattern.

To know individuals in which the match can be found, we have to identify all 
variants whose occurrence, or absence of,
have impact on the match, and list only the genomes with such combination of variants.






\subsection{Building the index}

To build the index, we process the input data to create the following main substructures, 
described in detail in the successive paragraphs:
\begin{itemize}
\item Reference sequence (REF),
\item Variant Database (VD),
\item Bit Vectors (BVs) with information about variants in all genomes,
\item The $k$-Mer Array ($k$MA) for all unique $k$-length sequences in the set.
\end{itemize}

REF is stored in compact form, where 4 bits are used to (conveniently) encode a single character. 

VD contains details about all possible variations. For each variant, the following items are stored: type (1 byte), preceding position\footnote{We keep the preceding positions to be able to manage the variants INSs, DELs and SVs, as this convention conforms to their description in VCF files.} (4 bytes) and alternative information (4 bytes). The last one indicates alternative character in case of SNP, length of the deletion in case of DEL and position in the additional arrays of bytes (VD-aux)  in case of INS and SV. VD-aux holds insertion length (4 bytes) and all inserted characters (1 byte each), if any, for every INS and SV. For SV it also stores length of the deletion (4 bytes).  The variants are ordered by the preceding position and a lookup table is created to 
accelerate search for a variant by its location.
VD together with REF can be used to decode the modified sequence from some given position to the right, by introducing certain variants. To be able to decode the sequence to the left, an additional list of all deletions (SVs and DELs), ordered by the resulting position, is created. The list, VD-invDel, stores for each variant its number in the main VD (4 bytes) and the resulting position, that is, the position in the reference after the deletion (4 bytes).

There is one BV for each variant, each of size of the number of genomes in the collection (2 times the number of genomes 
for 
diploid organisms). 
Value 1 at some $j$th position in this vector means that the current variant is found in the $j$th haploid genome.
To reduce the required size, while preserving random access, we keep the collection of these vectors in compressed form, 
making use of 
the fact that 
spatially close variant configurations are often shared across 
different individuals. 
The compression algorithm makes use of a dictionary of all possible unique 192-bit chunks (the size chosen experimentally). 
Each BV is thus represented as a concatenation of 
$\left \lceil{no\_haploid\_genomes/192}\right \rceil$ 
4-byte tokens (vocabulary IDs). 

To create $k$MA, we identify each $k$-mer occurring in the whole collection of genomes and keep minimum information to be able to retrieve it with help of REF and VD. 
For all $k$-mers in REF, the subarray $k$MA$^0$ is created, 
where only the preceding position \tuple{pos\_ref} ($4$ bytes) 
of each occurrence of the $k$-mer in REF is stored.
These $k$-mers are present in all genomes with no variants introduced in the corresponding segment. 
Based on the amount of details required for the $k$-mers 
to describe how they differ to a respective snippet of REF, 
we store them in one of the three subarrays: $k$MA$^1$, $k$MA$^2$ or $k$MA$^3$. 
These, together with $k$MA$^0$, form the complete $k$MA. 
 

To find all $k$-mers that differ from 
REF, we go through the reference genome and check for each position $p$ if there is any possible variant with the preceding position in the range from $p$ to $p+k-1$. If so, we decode the $k$-mer. The decoding process takes into account all paths, that is, all possible combinations of occurring variants. Thus, starting from a single preceding position, many resulting sequences may be obtained. 
To decode most $k$-mers, it is enough to store the preceding position plus flags about the occurrence/absence of neighboring variants. 
This evidence list ($\mathit{evList}$) is stored as a bit vector, where~$1$ means that the corresponding variant is found. For any $k$-mer starting inside an insertion (may be INS or SV) it is also necessary to store the $\mathit{gap}$ from the beginning of the inserted string to the first character of the $k$-mer.


The $k$-mer with no $\mathit{gap}$ and at most $32$ evidences about consecutive variants from the~VD in the $\mathit{evList}$ is stored in the $k$MA$^1$, where each 
entry is defined as \tuple{pos\_ref, evList} ($4+4$ bytes). If there is also a $\mathit{gap}$ involved, such $k$-mer goes to $k$MA$^2$, defining each 
entry 
as \tuple{pos\_ref, gap, evList} ($4+4+4$ bytes). All $k$-mers with more than $32$ evidences in the $\mathit{evList}$ or with evidences about nonconsecutive (with respect to the~VD) variants  are kept in $k$MA$^3$, where each $k$-mer is represented by four fields: \tuple{pos\_ref, gap, evSize, evList} ($4+4+4+\mathit{evSize}\times4$ bytes).  
The representative example of the latter case is a $k$-mer with SV introduced and many variants in the~VD placed within the deleted region. Keeping track of these variants, not alerting the resulting sequence, is pointless.

Any $k$-mer is kept in $k$MA only if there is at least one haploid genome that includes it, that is, has the same combination of occurring variants. It is checked with help of BV. 
The $k$-mers in each subarray $k$MA$^*$ are sorted alphabetically. To speed up the binary search (by narrowing down the initial search interval), a lookup table, taking into account the first 12 characters, is created for each subarray.

\subsection{The basic search algorithm}

The pseudocode of the basic search algorithm is given in 
Fig.~\ref{fig:pseudo:alg1}. It looks for all exact occurrences of the pattern $P$ in the compressed collection, using the seed-and-extend scheme.
The seed $S$ is chosen to be a substring of $P$, precisely its first $k$ characters, or the full pattern if $|P| < k$ (lines 1--2).

\begin{figure}[t]
\begin{myalgorithm}{The basic search algorithm (exact search for sequence $P$)}
\pccomment{search for seed S}\\
\>	\bfseries function \pcfun{locate}($P$)\\
\lno		$p \gets \pcfun{min}(|P|, k)$\\
\lno		$S \gets \pcfun{substring}(P, 0, p-1)$\\
\lno		\pcfor $i \gets 0$ \pcto $3$ \pcdo\\
\lno\>		$(l, r) \gets \pcfun{binSearch}(k$MA$^i, S)$\\
\lno\>		\pcfor $j \gets \ell$ \pcto $r$ \pcdo\\
\lno\>\>		$\mathit{vtList.\pcfun{reset}}(); \mathit{evList.\pcfun{reset}}()$\\
\lno\>\>		$(\mathit{vtList}, \mathit{evList}, \mathit{pos}, \mathit{vt}) \gets \pcfun{partDecode}(k$MA$^i[j], p)$\\
\lno\>\>		$\pcfun{extend}(k$MA$^i[j].\mathit{pos\_ref}, \mathit{pos}, p, \mathit{vt})$\\
\pccomment{function extending the found seed}\\
\>	 \bfseries function \pcfun{extend}$(pre, \mathit{pos}, \mathit{ch}, \mathit{vt})$\\
\lno		\pcwhile $\mathit{ch} < \mathit{P.len}$ \pcdo\\
\lno\>		\pcif $\mathit{vt.pos} > \mathit{pos}$ \pcthen\pccomment{No variant at pos}\\
\lno\>\>	 	\pcif $\mathit{REF}[\mathit{pos}]$ matches $P$ \pcthen\\
\lno\>\>\>		$\mathit{pos} \gets \mathit{pos} + 1; \mathit{ch} \gets \mathit{ch} + 1$\\
\lno\>\>		\pcelse report \pcfalse \pccomment{Wrong path}\\
\lno\>		\pcelseif  $\mathit{vt.pos} = \mathit{pos}$ \pcthen\\
\lno\>\>		$\mathit{vtList.\pcfun{add}(vt)}; \mathit{evList.\pcfun{add}(1)};$\\
\lno\>\>		\pcif $\mathit{vt}$ matches $P$ \pcthen\\
\lno\>\>\>		$ \mathit{new} \gets \mathit{pos} + \mathit{vt.delLen}$\\
\lno\>\>\>		$\pcfun{extend}(pre, \mathit{new}, \mathit{ch }+ \mathit{vt.len}, \mathit{vt}+1)$\\
\lno\>\>		$\mathit{evList.\pcfun{setLast}(0)}; \mathit{vt} \gets \mathit{vt} + 1$\\
\lno\>		\pcelse \pccomment{$\mathit{vt.pos} < \mathit{pos}$}\\
\lno\>\>		$\mathit{new} \gets \mathit{vt.pos} + \mathit{vt.delLen}$\\
\lno\>\>		\pcif $\mathit{new} > \mathit{pos}$ \pcthen\\
\lno\>\>\>		$\mathit{vtList.\pcfun{add}(vt)}; \mathit{evList.\pcfun{add}(1)};$\\
\lno\>\>\>		\pcif $\mathit{vt}$ matches $P$ \pcthen\\
\lno\>\>\>\>		$\pcfun{extend}(pre, \mathit{new}, \mathit{ch} + \mathit{vt.len},\mathit{vt}+1)$\\
\lno\>\>\>		$\mathit{evList.\pcfun{setLast}(0)}$\\
\lno\>\>		$\mathit{vt} \gets \mathit{vt} + 1$\\

\lno		$R \gets 1^{\mathit{noHaploidGenomes}}$\\
\lno		\pcfor $i \gets 1$ \pcto $\mathit{vtList.size}$ \pcdo\\
\lno\>		\pcif $\mathit{evList}[i]$ \pcthen $R \gets R  ~\band \mathit{BV}[i]$\\
\lno\>		\pcelse $R \gets R  ~\band \bneg \mathit{BV}[i]$\\
\lno		\pcif $R = 0$ \pcthen report \pcfalse \pccomment{Wrong path}\\
\lno		\pcelse report $(pre, R)$ \pccomment{P found}
\end{myalgorithm}
\caption{The basic searching algorithm}
\label{fig:pseudo:alg1}
\end{figure}

The first step is to scan the~$k$MA for all $k$-mers matching~$S$. It is done with binary search in each subarray $k$MA$^*$ separately (lines 3--4). Next, each found seed is partly decoded (only number of decoded characters is counted) and then extended (lines 5--8). The decoding is based on the~$k$-mer's details to get the seed's succeeding position ($\mathit{pos}$) and variant ($\mathit{vt}$) in the reference, along with the list of encountered variants ($\mathit{vtList}$) and the list of evidences about their occurrence or absence of ($\mathit{evList}$). The latter is a vector of $0$s in case of $k$MA$^0$ and a copy of $k$-mer's $\mathit{evList}$ (or its part) for other subarrays. The first variant (the one with preceding position greater than or equal to the preceding position of the $k$-mer) is found with binary search in VD. It is not shown in the pseudocode, but for each seed also the preceding SVs and DELs are taken into account when creating the initial $\mathit{vtList}$ and  $\mathit{evList}$.

The seed $S$ is recursively extended according to all possible paths, that is as long as succeeding characters match the characters in $P$ (lines 9--33). Maintained variables are:  $s$ and $pos$ (the preceding position of the seed and the current position, both in relation to the reference),  $ch$ (number of decoded characters) and $\mathit{vt}$ (next variant from VD). Also the current $\mathit{vtList}$ and $\mathit{evList}$ are available. 
If position of $\mathit{vt}$ ($\mathit{vt.pos}$) is greater than $pos$ (lines 10--13), no variant is introduced and the next character is taken from REF. If it does not match the related character in $P$, the extension is stopped, as the current path is not valid. 
If $\mathit{vt}$ is encountered at $pos$ (lines 14--19), it is added to the $\mathit{vtList}$ and two paths are checked---when it is introduced (new bit in $\mathit{evList}$ is set to $1$) and when it is not (new bit in $\mathit{evList}$ is set to $0$). The first path is not taken if $\mathit{vt}$ does not match~$P$. It can happen for SNPs and inserted characters (from INS or SV).
If $\mathit{vt.pos}$ is less than  $pos$ (lines 20--27), it means $\mathit{vt}$ is placed in region previously deleted by other variant. The only possibility that  $\mathit{vt}$  is taken into account is if it deletes characters beyond previous deletion. Otherwise it is skipped.

When the extension reaches the end of the pattern $P$, it is checked in which individuals, if in any, the relevant combination of significant variants (track kept in $\mathit{vtList}$) is found (lines 28--33). The bit vector $R$ is initialized to be the size of number of haploid genomes. The value $1$ at $j$th position means that $j$th haploid genome contains the found sequence. 
The vector $R$ is set to all $1$s at the beginning, because if $\mathit{vtList}$ is empty, the sequence is present in all genomes.
To check which genomes have the appropriate combination of variants, the bitwise AND operations are performed between all BVs related to variants from the $\mathit{vtList}$, negating all BVs with $0s$ at the corresponding position in the $\mathit{evList}$.
If $R$ 
contains any 1s,
pattern $P$ is reported to be found with the preceding position $pre$ (in relation to the reference genome) and vector~$R$ specifies genomes containing such sequence.

\subsection{The space-efficient version}

To reduce the required space, while still being able to find all
occurrences of the pattern, we make use of the idea of sparse suffix array~\citep{KU1996}.
This data structure stores only
the suffixes with preceding position being a multiplication of $s$ 
($s > 1$ is a construction-time parameter).
In our scheme, the two largest subarrays, $k$MA$^0$ and $k$MA$^1$,
are kept in sparse form, based on preceding positions of $k$-mers.
For $k$MA$^1$, it is also necessary to remain all $k$-mers that begin with deletion or insertion (the first variant has the same preceding position as the $k$-mer). 

The search algorithm has to be slightly modified. 
Apart from looking for the $k$-length prefix of the pattern 
(i.e., $P[0\ldots k-1]$)
in $k$MA, also $k$-length substrings starting at positions 
$1 \ldots s-1$
must be looked for in  $k$MA$^0$,  $k$MA$^1$, and $k$MA$^3$ (as some specific seeds may be present only in $k$MA$^3$). 
The substrings, if found in one of mentioned subarrays, must be then decoded to the left, to check if their prefix (from $1$ to $s-1$ characters, depending on the starting position) 
matches the pattern~$P$.
The VD-invDel substructure is 
used 
for the process.
The rest of search is the same as in the basic search algorithm.

\begin{table*}[t]
\processtable{Query times for various variants of indexes\label{tab:times}}{
\renewcommand{\tabcolsep}{1.5em}
\begin{tabular}{cccccccccc}\toprule
$k$	& Sparsity		& Size  		&&	\multicolumn{6}{c}{Max.\ allowed mismatches}\\\cline{5-10}
		&		& [GB]	&&	0			& 1			& 2			& 3			& 4			& 5			\\\midrule
	40	& \q1 & 29.6	&&	\q28.8	& \q65.2		& 102.3		& \qc291.0	& 1,109.9	& 3,348.8	\\
	40	& \q3	& 13.4	&&	\q39.4	& \q85.5		& 136.1		& \qc372.5	& 1,334.0	& 4,021.0	\\
	40	& \q4	& 11.4	&&	\q43.4	& \q94.4		& 151.4		& \qc412.2	& 1,471.1	& 4,461.1	\\
	40	& \q8	& \q8.4	&&	\q61.0	& 128.9		& 210.3		& \qc615.4	& 2,297.9	& 7,350.3	\\
	40	& 12	& \q7.4	&& \q76.3	& 160.0		& 271.8		& \qc917.0	& --- & ---\\
	40	& 16	& \q6.9	&&	\q90.4	& 184.4		& 344.3		& 1,514.1  	& --- & ---\\
\midrule
\multicolumn{2}{c}{GEM mapper}	
				& \q5.0	&& \q14.3		& \q26.6		& \q40.4		& \qq71.9	& \qc126.7	& \qc262.7	\\
\bottomrule
\end{tabular}}
{All times are expressed in $\mu$s. 
We do not provide times for large sparsities and more errors than 3, since in such cases the internal queries would be for very short sequences and in turn result in numerous matches and significant times; thus, we do not recommend to use MuGI in such configurations of parameters}
\end{table*}

\subsection{The approximate search algorithm}

The approximate search algorithm looks for all occurrences of the given pattern with some maximum allowed number of mismatches.
For any sequence of length $\ell$ with $m$ mismatches at least one of the consecutive substrings of length  $q = \left \lfloor{\frac{\ell}{m+1}}\right \rfloor$ is the same as in the original sequence. Therefore, the approximate search begins with dividing the string to $m+1$ substrings of length~$q$. Next, the exact search algorithm is used to look for each of the substrings. If a substring is found in the collection, it is further decoded to the right and to the left, similarly as in the exact search, but 
allowing for 
at most $m$ differences between the decoded sequence and the searched sequence. Expanding to the left is done with aid of the same auxiliary substructure as 
in the 
space-efficient version (VD-invDel). The list of genomes in which the found sequences are present is obtained in the same way as in the exact searching.

\subsection{Test data}
To evaluate the algorithm, we used a similar methodology as the one 
in~\citep{WSBL2013}.
To this end, we generated a file with 100K queries, where 
each pattern is a modified excerpt of length $\ell=120 \ldots 170$ (uniformly random value) from a randomly selected genome from the collection, starting at a randomly selected position.
Excerpts containing unknown characters (i.e., N) were rejected. The modifications consisted in introducing random nucleotides in place of $x$ existing nucleotides, where $x$ is a randomly selected integer number from the $[0, 0.05\times\ell)$ range.

\section{Results}

All experiments were performed on a PC with Intel i7 4770 3.4\,GHz CPU 
(4 cores with hyperthreading), equipped with 32\,GB of RAM, running Windows 7 OS.
The C++ sources were compiled using GCC 4.7.1 compiler.

\begin{figure*}
\ifpdf
\centering\includegraphics[bb=66 537 542 708, clip]{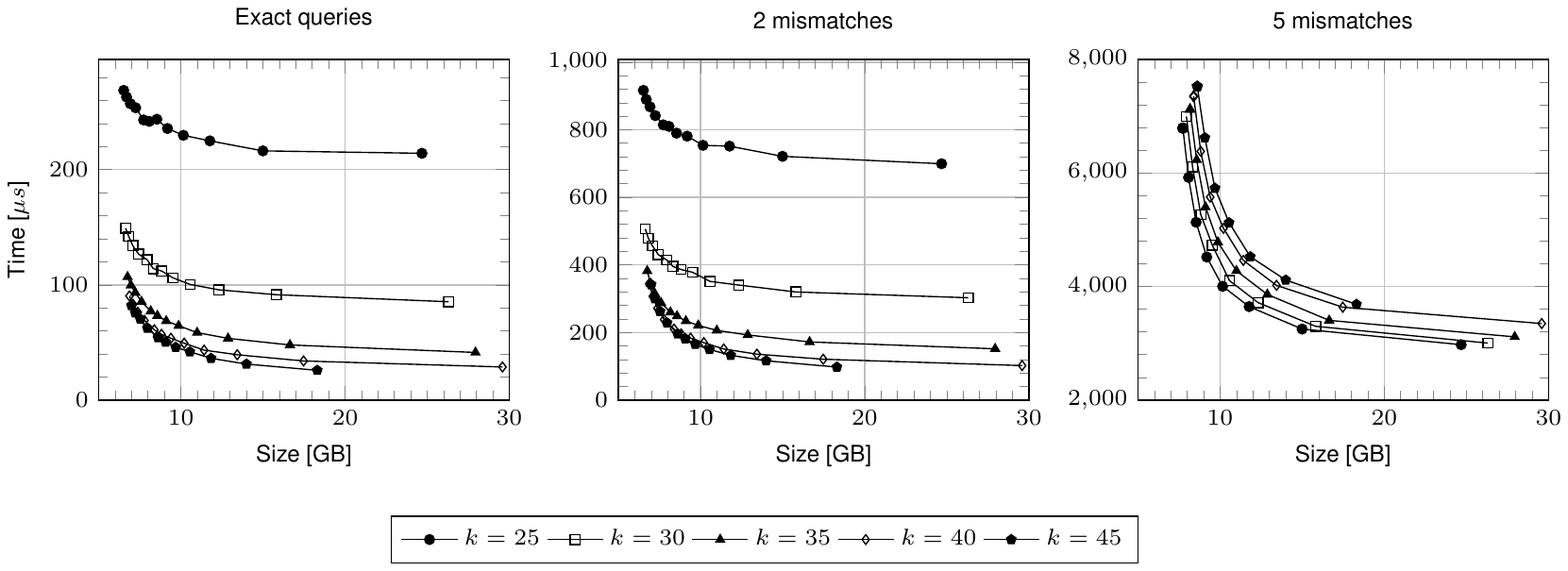}
\else
\centering\includegraphics[bb=66 537 542 708, clip]{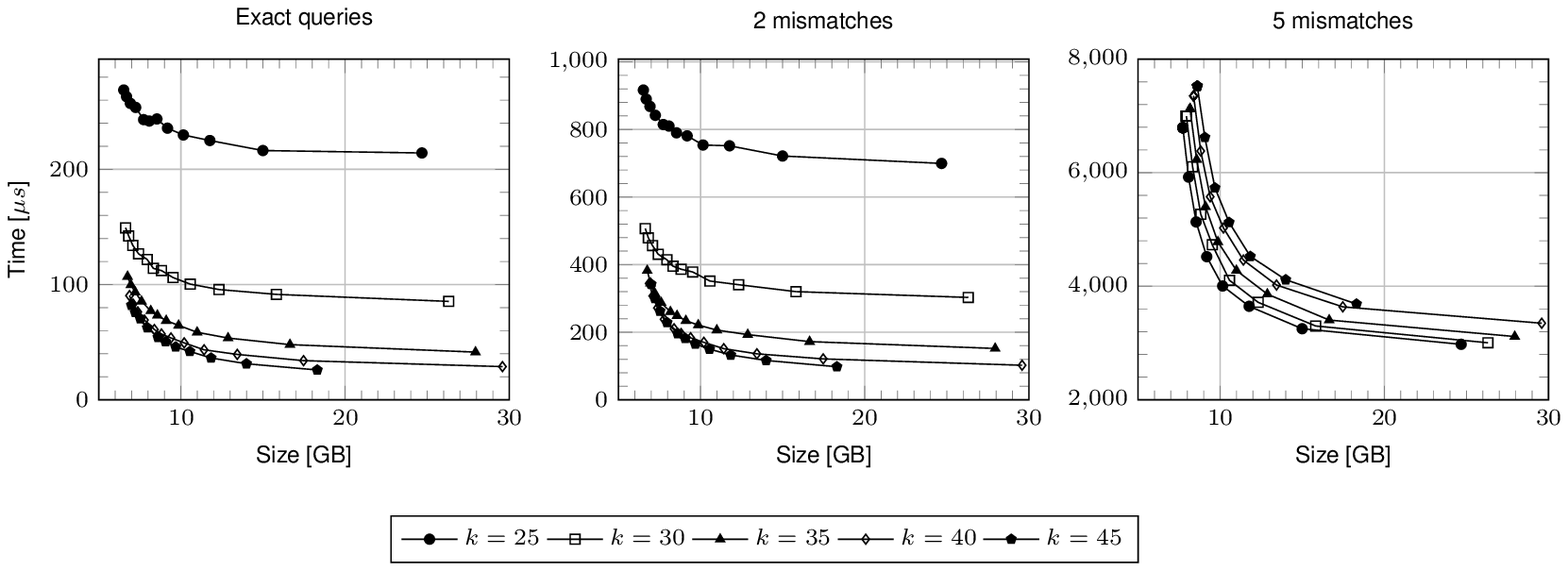}
\fi
\caption{Average query times vs. index sizes}
\label{fig:locate-times}
\end{figure*}

The index was built on another machine (2.4\,GHz Quad-Core AMD Opteron CPU
with 128\,GB RAM running Red Hat 4.1.2-46) and required more RAM: 
from 38\,GB (for $k=25$) to 47\,GB (for $k=45$). 
The corresponding build times were 15 hours and 72 hours, respectively.
The index build phase was based on parallel sort (using Intel TBB and OpenMP libraries), while all the queries in our experiments were single-threaded. 

From Table~\ref{tab:index-sizes} 
we can see that the fastest index version 
(i.e., with sparsity 1, which translates to 
standard $k$-mer arrays)
may work 
on the test machine even for the seed maximum length of 40 symbols.
Significant savings in the index size are however possible if sparsity of 3 
or more is set, making the index possible to operate on a commodity PC with 
16\,GB of RAM.
If one (e.g., a laptop user) requires even less memory, then 
the sparsity set to 16 makes it possible to run the index even in 8\,GB of RAM.
Naturally, using larger sparsities comes at a price of slower searches;
in Fig.~\ref{fig:locate-times}, each series of results for a given value of $k$ 
corresponds to sparsities from $\{1, 2, \ldots, 8, 10, 12, 14, 16\}$ 
(sparsities of 1 correspond to the rightmost points, with the exception 
of the case of $k=45$, for which the sparsities start from 2).
Still, this tradeoff is not very painful: 
even the largest allowed sparsity value (16) slows down the fastest 
(for sparsity of 1) queries by factor about 2 on average, in most cases.

\begin{table}[t]
\processtable{Index sizes\label{tab:index-sizes}}{
\renewcommand{\tabcolsep}{0.75em}
\begin{tabular}{cccccc}\toprule
Sparsity	&	\multicolumn{5}{c}{Size [GB]}\\\cline{2-6}
		& $k=25$	& $k=30$	& $k=35$	& $k=40$	& $k=45$	\\\midrule
\q1	& 24.7	& 26.3	& 27.9	& 29.6	& 31.2	\\
\q2	& 15.0	& 15.8	& 16.6	& 17.5	& 18.3	\\
\q3	& 11.8	& 12.3	& 12.9	& 13.4	& 14.0	\\
\q4	& 10.2	& 10.6	& 11.0	& 11.4	& 11.8	\\
\q5	& \q9.2	& \q9.5	& \q9.9	& 10.2	& 10.5	\\
\q6	& \q8.5	& \q8.8	& \q9.1	& \q9.4	& \q9.7	\\
\q7	& \q8.1	& \q8.3	& \q8.6	& \q8.8	& \q9.1	\\
\q8	& \q7.7	& \q7.9	& \q8.2	& \q8.4	& \q8.6	\\
10		& \q7.2		& \q7.4	& \q7.6	& \q7.8	& \q8.0	\\
12		& \q6.9		& \q7.1	& \q7.2	& \q7.4	& \q7.5	\\
14		& \q6.7		& \q6.8	& \q7.0	& \q7.1	& \q7.2	\\
16		& \q6.5		& \q6.6	& \q6.8	& \q6.9	& \q7.0	\\
\bottomrule
\end{tabular}}
{}
\end{table}

Costlier, in terms of query times, is handling mismatches. 
In particular, allowing 4 or 5 mismatches in the pattern requires at least 
an order of magnitude longer query times than in the exact matching mode.
Yet, even for 5 allowed errors the average query time was below 10\,ms in all tests.

Apart from the average case, one is often interested also in the pessimistic 
scenario.
Our search algorithms do not have interesting worst-case time complexities, 
but fortunately pathological cases are rather rare.
To measure this, for each test scenario a histogram of query times over 100K patterns 
was gathered, and the time percentiles are shown in 
Fig.~\ref{fig:percentile-times}.
Note that the easy cases dominate: for all maximum errors allowed, 
for 90\% test patterns the query time is below the average.
Yet, there are a few percent of test patterns for which the times are several 
times longer, 
and even a fraction of a percent of patterns with query times exceeding 100\,ms 
(at least for approximate matching).
More details exposing the same phenomenon are presented 
in Table~1 in the Supplementary Material.

\begin{figure}
\ifpdf
\includegraphics[bb=2 650 229 840, clip]{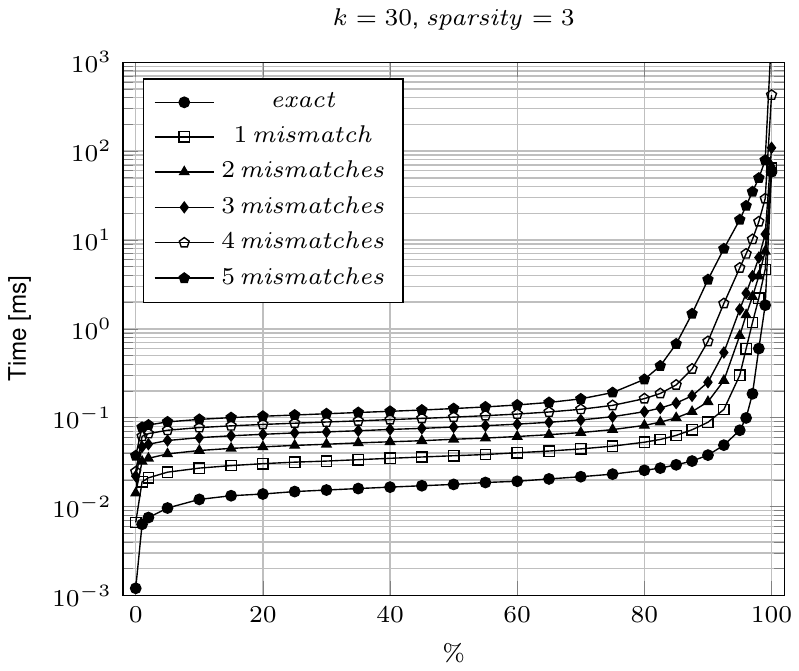}
\else
\includegraphics[bb=2 650 229 840, clip]{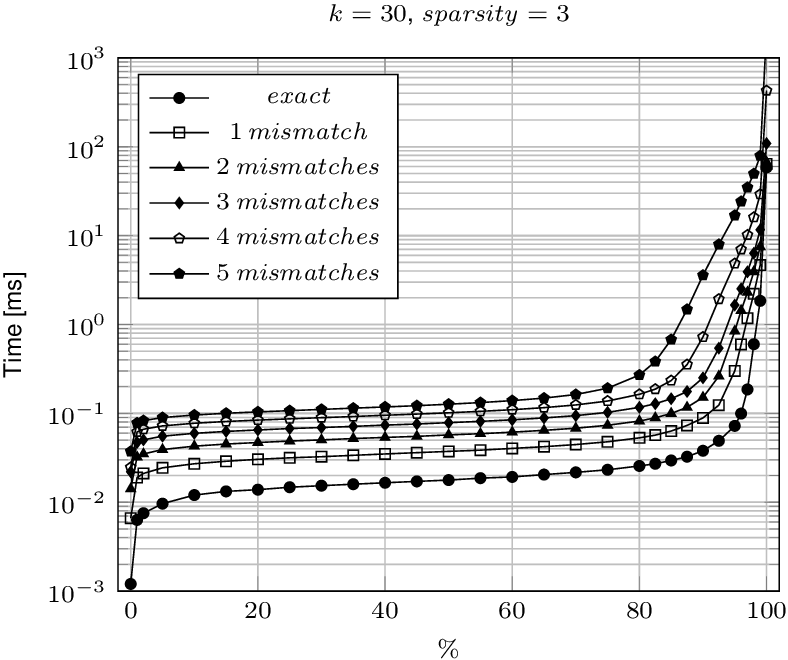}
\fi
\caption{Query time percentiles for exact and approximate matching, for max error up to 5. For example, the 80th percentile for 1 error equal to 0.52\,ms means that 80\% of the test patterns were handled in time up to 0.52\,ms each, allowing for 1 mismatch.}
\label{fig:percentile-times}
\end{figure}

While we cannot directly compare our solution to RCSI by 
Wandelt {\it et~al.}~(\citeyear{WSBL2013}), as their software wasn't available to us, 
we can show some comparison.
Their index was built over twice less data 
(haploid human genomes vs. diploid genomes in our data).
We handle exact matches much faster (over 6 times shorter reported average times, 
but considering the difference in test computers this probably translates 
to factor about 4).
Roughly similar differences can be observed for the approximate matching scenario, 
but RCSI handles the Levenshtein distance ($k$-differences), while our scheme 
handles (so far) only mismatches.
Finally, and perhaps most importantly, our index over 1092 diploid human genomes 
can be run on a standard PC, equipped with 32 or 16\,GB of RAM 
(or even 8\,GB, for the price of more slow-down), while RCSI requires a machine 
with 128\,GB (unless searches are limited to one chromosome, when a portion of 
the index may be loaded into memory).\looseness=-1

We did, however, ran a preliminary comparison of MuGI against GEM~\citep{MSSGR2012}, 
one of the fastest single genome read mappers.
We ran it on 1 CPU core, for mismatches only, in the all-strata mode, 
in which all matches with $0, 1, \ldots, max\_mismatches$ errors are reported, 
in arbitrary order.
Table~\ref{tab:times} contains a brief comparison 
(for a detailed rundown see Table~2 in the Supplementary Material).
For example, we can see that GEM performed  
exact matching in 14.3$\mu$s,
found matches with up to 1 mismatch in 26.6$\mu$s,
matches with up to 3 mismatches in 71.9$\mu$s,
matches with up to 5 mismatches in 262.7$\mu$s.
The memory use was 5.0\,GB.
This means that, depending on chosen options of our solution, 
GEM was only about 2--3 times faster in the exact matching 
mode and 13--15 times faster when 5 mismatches were allowed.
The major scenario difference is however that GEM performs mapping to a single 
(i.e., our reference) genome, so to obtain the same mapping results 
GEM would have to be run $2 \times 1092$ times, once per haploid genome. 
We thus consider these preliminary comparative results very promising.

\end{methods}

\section{Conclusions and future work}

We presented an efficient index for exact and approximate searching over large repetitive genomic collections, in particular: multiple genomes of the same species.
This has a natural application in aligning sequencing reads against a collection of genomes, with expected benefits for, e.g., personalized medicine and deeper understanding of the interaction between genotype and phenotype.
Experiments show that the index built over a collection of $2 \times 1092$ human genomes fits a PC machine with 16\,GB of RAM, or even half less, for the price of some slow-down.
According to our knowledge, this is the first feat of this kind.
The obtained solution is capable of finding all pattern occurrences in the collection in much below 1\,ms in most use scenarios.

Several aspects of the index require further development. 
The current approximate matching model comprises mismatches only; 
it is desirable to extend it to edit distance.
The pathological query times could be improved with extra heuristics 
(even if it is almost irrelevant for large bulk queries).
A more practical speedup idea is to enhance the implementation with 
multi-threading.
Some tradeoffs in component data structures 
(cf. Table~3 in the Supplementary Material) 
may be explored, e.g., 
the reference genome may be encoded more compactly but at a cost of somewhat 
slower access.
A soft spot of the current implementation is the index construction phase, 
which is rather na\"ive and can be optimized especially towards reduced memory 
requirements.
We believe that existing disk-based suffix array creation algorithms 
(e.g.,~\citep{K2007}) can be adapted for this purpose.
Alternatively, we could build our indexing data structure separately 
for each chromosome (with memory use for the construction reduced by 
an order of magnitude) and then merge those substructures, onto disk, 
using little memory.
Finally, experiments on other collections should be interesting, 
particularly on highly-polymorphic ones.

\section*{Acknowledgement}

\paragraph{Funding\textcolon}
The work was supported by the Polish Ministry of Science and Higher Education under the
project DEC-2013/09/B/ST6/03117.
The work was performed using the infrastructure supported by POIG.02.03.01-24-099/13 grant: ``GeCONiI---Upper Silesian Center for Computational Science and Engineering''.

\subsection*{Conflict of interest statement.}
None declared.


\end{document}